\documentclass[%
 aip,
 amsmath,amssymb,
 reprint,%
]{revtex4-1}

\usepackage{graphicx}
\usepackage{dcolumn}
\usepackage{bm}

\usepackage[utf8]{inputenc}
\usepackage[T1]{fontenc}
\usepackage{mathptmx}

\begin{document}

\preprint{AIP/123-QED}

\title[Synchronization Behavior in a Ternary Phase Model]{Synchronization Behavior in a Ternary Phase Model}

\author{N. DeTal}
 \email{ndetal3@gatech.edu}
 \affiliation{ 
Center for Nonlinear Science, School of Physics, Georgia Institute of Technology, Atlanta, Georgia 30332, USA
}%
\author{H. Taheri}%
\email{hossein.taheri@ucr.edu}
\affiliation{ 
Department of Electrical and Computer Engineering, University of California at Riverside, 900 University Ave., Riverside, California 92521, USA
}%

\author{K. Wiesenfeld}
\affiliation{ 
Center for Nonlinear Science, School of Physics, Georgia Institute of Technology, Atlanta, Georgia 30332, USA
}%

\date{\today}

\begin{abstract}
Localized traveling-wave solutions to a nonlinear Schr\"odinger equation were recently shown to be a consequence of Fourier mode synchronization. The reduced dynamics describing mode interaction take the form of a phase model with novel ternary coupling. We analyze this model in the presence of quenched disorder and explore transitions to partial and complete synchronization. For both Gaussian and uniform disorder, first-order transitions with hysteresis are observed. These results are compared with the phenomenology of the Kuramoto model which exhibits starkly different behavior. An infinite-oscillator limit of the model is derived and solved to provide theoretical  predictions for the observed transitions. Treatment of the nonlocal ternary coupling in this limit sheds some light on the model's novel structure.
\end{abstract}

\maketitle

\begin{quotation}
The damped, driven nonlinear Schr\"odinger equation (NLSE) with cubic nonlinearity is ubiquitous in nonlinear science. This equation models a broad range of physical systems, for instance charge-density waves, Josephson junctions, ferromagnets in microwave fields, quantum Hall ferromagnets, radio-frequency-driven plasmas, shear flows in liquid crystals, ocean and atmospheric waves, and spatial and temporal nonlinear waves in optical resonators with Kerr nonlinearity. These nonlinear wave phenomena can all exhibit sharply-peaked patterns of spatially and/or temporally localized pulses. In these systems, pulsation can be a signature of synchronization between many oscillating degrees of freedom. A previously derived model\cite{taheri} explicitly relates the existence of such a pulse to the synchronization of its Fourier modes. Here, we analyze this model in the context of general coupled-oscillator systems and investigate its synchronization behavior. 
\end{quotation}

\section{Introduction}

Mode locking, i.e. alignment of laser mode frequencies on an equidistant spectral grid, has long been known to be a requirement for pulsation in actively and passively mode-locked lasers \cite{mcduff,  kuizenga, sargent}. Variants of the nonlinear Schr\"odinger equation (NLSE) that admit sharply peaked soliton pulses have successfully been used to model many types of mode locked lasers \cite{haussat, hausstruc, hausloc, martinez}. The connection between mode locking (pulsation in the time domain) and the universal phenomenon of self-synchronization is evidenced by the alignment and synchronous oscillation of Fourier modes in the frequency domain, see, e.g., (\onlinecite{haken, wen, taheri2}). Recently, this nexus was formally established through a phase model underlying spontaneous pulse formation in the damped and driven NLSE \cite{taheri}, also called the Lugiato-Lefever equation (LLE) \cite{lugiato}.

Interestingly, this phase model is of a novel type within the realm of self-synchronization phenomena: the phases interact in triples rather than pairs.  The ternary coupling is rooted in the physics of the nonlinear interaction; specifically, the conservation of energy and momentum in nonlinear four-wave mixing in optical Kerr resonators. In this work, we treat the ternary phase model as a stand-alone model worthy of independent investigation from a nonlinear dynamical perspective.  In particular, we explore the transitions to partial and complete synchronization, and compare the basic phenomenology to that of the Kuramoto model, the archetype of pairwise-interacting oscillator populations. 

The paper is organized as follows.  In Section II we introduce the ternary phase model and briefly review its relationship to the NLSE.  In Section III, we present the results of numerical simulations showing the transition behavior for different choices of disorder, and compare this with the corresponding behavior of the Kuramoto model. In Section IV we construct a ``thermodynamic limit'' of the model and derive an expression we then use to calculate the order parameter quantifying synchronization. We end with a discussion of open questions. 

\section{Background}
In this paper, we explore a disordered version of the following coupled oscillator equation \cite{taheri}

\begin{equation}
    \begin{split}
         \dot{\Delta}_j&=\frac{K}{2N} \sum_{l=j-N}^{N} \sin(\Delta_l+\Delta_{j-l}-\Delta_j), \qquad ; \qquad j = 1, 2, \dots, N \\
    \end{split}
    \label{ternarymodel}
\end{equation}
where $\Delta_j$ is the phase of the $j^{th}$ oscillator, and $K$ is the coupling constant. By definition, $\Delta_{-j} = -\Delta_j$. The equation bears some resemblance to the Kuramoto model \cite{kuramoto}.  We briefly review how Eq.(\ref{ternarymodel}) comes about; full technical details can be found in Ref. (\onlinecite{taheri}).  

Equation \ref{ternarymodel} governs the slow dynamics of the LLE in the strongly pumped limit.  As first noted by Wen et al, in this limit, the dynamical evolution of the system proceeds in three successive steps, which can be captured by a systematic expansion in inverse powers of the pump amplitude\cite{wen}.  On the fastest time scale, the pumped mode steady-state is established.  On the intermediate time scale, mode pairs (symmetrically located about the pump frequency) grow via parametric amplification.  Each mode pair evolves independently, reaching its steady state amplitude, and simultaneously a partial ordering of its phases such that $\phi_{j} + \phi_{-j} - 2\phi_0 = 0$. (Indices are chosen such that the pump index is zero.) It is during this intermediate time scale that a dense frequency comb is established; however, localized/pulsing dynamics emerges only on a still slower time scale, described by Eq.(\ref{ternarymodel}), during which the anti-symmetric parts of the phase pairs $\Delta_j = \phi_{j} - \phi_{-j}$ interact with each other. In the context of the LLE, $\Delta_j$ specifies a phase difference between modes. In this paper, we consider the set to be absolute phases of an oscillator population. The three-phase coupling of (\ref{ternarymodel}) is the direct consequence of the cubic nonlinearity in the underlying wave equation.  We note that phase models with three- and higher-fold couplings are not unheard of, and in fact arise generically in certain contexts \cite{ashwin}.  

Our first observation is that Eq.(\ref{ternarymodel}) admits a family of perfectly synchronized states, given by

\begin{equation}
    \Delta_j=\alpha \cdot j
    \label{ordered_state}
\end{equation} 
for any constant $\alpha$.  In the context of the micro-resonator problem, these states correspond to a spatially localized pulse propagating around the micro-resonator ring, and $\alpha$ determines the location of the optical pulse in the co-moving frame. This definition of synchronization is analogous to the situation in the Kuramoto model, in which the oscillators converge to an arbitrary phase
\begin{equation}
    \theta_j = \beta.
\end{equation}
For the ternary model, the oscillators will have identical phases if $\alpha=0$, though there is nothing unique about that particular value. 

 It has been shown \cite{taheri} that this state is dynamically stable except for a single neutral eigenvalue corresponding to a shift in $\alpha$.

In what follows, we consider a generalization of Eq.(\ref{ternarymodel}), where we include some intrinsic quenched disorder:

\begin{equation}
\dot{\Delta}_j=\omega_j+\frac{K}{2N}\sum_{l=j-N}^N \sin(\Delta_l+\Delta_{j-l}-\Delta_j)
\label{ternary_with_disorder}
\end{equation}
with the disorder constants $\omega_j$ drawn from a fixed, zero-mean distribution $g(\omega)$. We expect such imperfections on physical grounds.  In the optical resonator context, the $\omega_j$ arise from cavity imperfections and/or high-order dispersion effects.

\section{Synchronization Behavior}
We now consider the collective behavior of the ternary phase model, in particular the tendency to evolve into ordered states.  Our approach is inspired by what is known about the Kuramoto model.  To begin, we note that in the case of zero disorder ($\omega_j = 0$), Eq.(\ref{ternary_with_disorder}) is a gradient system $\dot{\Delta}_j = -\partial V/\partial \Delta_j$
with potential given by
\begin{equation}
        V=-\frac{K}{6N}\sum_{j=1}^N \sum_{l=j-N}^N \cos (\Delta_l+\Delta_{j-l}-\Delta_j).
\end{equation}
This structure implies the existence of locally stable perfectly synchronized states \cite{gushchin}.  Extensive numerical simulations suggest that, for almost all initial conditions, the disorder-free system evolves to a perfectly synchronized state.  (In simulations of Eq.(\ref{ternarymodel}), a state satisfying condition (\ref{ordered_state}) may not be immediately obvious, owing to the angular nature of the variables.  A simple check can be performed by plotting the difference of adjacent $\Delta_j$ {\it modulo} $2\pi$.) In view of this, we expect the dynamics of Eq.(\ref{ternary_with_disorder}) to exhibit a competition between the synchronizing influence of the coupling and the de-synchronizing influence of the disorder.  The central issue is whether, and under what circumstances, the system exhibits an order-disorder transition, and if so what is the nature of the transition(s).

In order to investigate this question quantitatively, we introduce the following set of complex order parameters:
\begin{equation}
    Z_j=\frac{1}{2N+1-j}\sum_{l=j-N}^{N}\textrm{e}^{\textrm{i}(\Delta_{l}+\Delta_{j-l})}.
    \label{orderp}
\end{equation}
$j=1, 2, \dots, N$.  At steady state, $|Z_j|$ is observed to be independent of index $j$, up to fluctuations of order $N^{-1/2}$. For a perfectly synchronized state (\ref{ordered_state}),
$|Z_j|$ is exactly one, while for randomly distributed $\Delta_j$ it tends to zero. In practice, it is useful to average $|Z_j|$ over all oscillators and consider $\overline{|Z|}$ to be the overall order parameter.

Our simulations show that, for long times, trajectories of Eq.(\ref{ternary_with_disorder}) evolve linearly in time for all $\Delta_j$, {\it i.e.}
\begin{equation}
    \Delta_j \rightarrow \alpha \cdot j+ \delta_j+\Omega_j \cdot t .
    \label{tsync1}
\end{equation}
The $\delta_j$ reflect deviation from (\ref{ordered_state}) due to the added disorder. From Eq.(\ref{ternary_with_disorder}), this relationship can only be satisfied if 
\begin{equation}
    \Omega_j=\Omega \cdot j .
    \label{tsync2}
\end{equation}
Taking a weighted sum of the $N$ equations yields, since the coupling terms cancel,
 
\begin{equation}
 \sum_{j=1}^N  \dot{\Delta}_j \cdot j= \sum_{j=1}^N \omega_j \cdot j 
\end{equation}
so that
\begin{equation}
\sum_{j=1}^N (\Omega \cdot j) \cdot j= \sum_{j=1}^N \omega_j \cdot j
\end{equation}
and so
\begin{equation}
\Omega = \frac{6}{N(N+1)(2N+1)} \cdot \sum_{j=1}^N \omega_j \cdot j .
\label{omega}
\end{equation}
For convenience, we can consider the co-rotating variables
\begin{equation}
\Delta'_j = \Delta_j - \Omega \cdot j \cdot t.
\label{rotframe}
\end{equation}
From (\ref{tsync1}) and (\ref{tsync2}), we can easily see that the transformed system $\Delta'_j$ evolves to a stationary state. Incidentally, this transformation also ensures the existence of a constant of motion
\begin{equation}
    \Gamma=\sum_{j=1}^N \Delta'_j \cdot j .
\end{equation}
Unless otherwise noted, the simulations described below are performed in this frame by generating a set of $\omega_j$, calculating $\Omega$, and subtracting according to Eq.(\ref{rotframe}).

To investigate order-disorder transitions in our system, we performed simulations using three different distributions for the $\omega_j$: uniform random, Gaussian random, and a systematic deterministic disorder to be described later. 

The simulations were carried out as follows.  We started by generating a random set of $\omega_j$ with unit variance. Multiplying this set by a factor $\sigma$ gives a standard deviation $\sigma$ while retaining comparable statistical properties. Starting with $\sigma=0$ and random initial conditions, we numerically integrated Eq. (\ref{ternary_with_disorder}) until the system reached its stationary synchronized state. We then increased $\sigma$ by a small amount and continued integrating, until a new steady state was reached (identified by an unchanging averaged order parameter $\overline{|Z|}$). Iterating this process, the system eventually reaches the unsynchronized regime at $\sigma_1$. At some distance into the unsynchronized regime we reversed the process by gradually reducing $\sigma$.  Eventually, at $\sigma_2$, the system begins to synchronize again.

\begin{figure*}[h!]
\centering
\includegraphics[scale=.4]{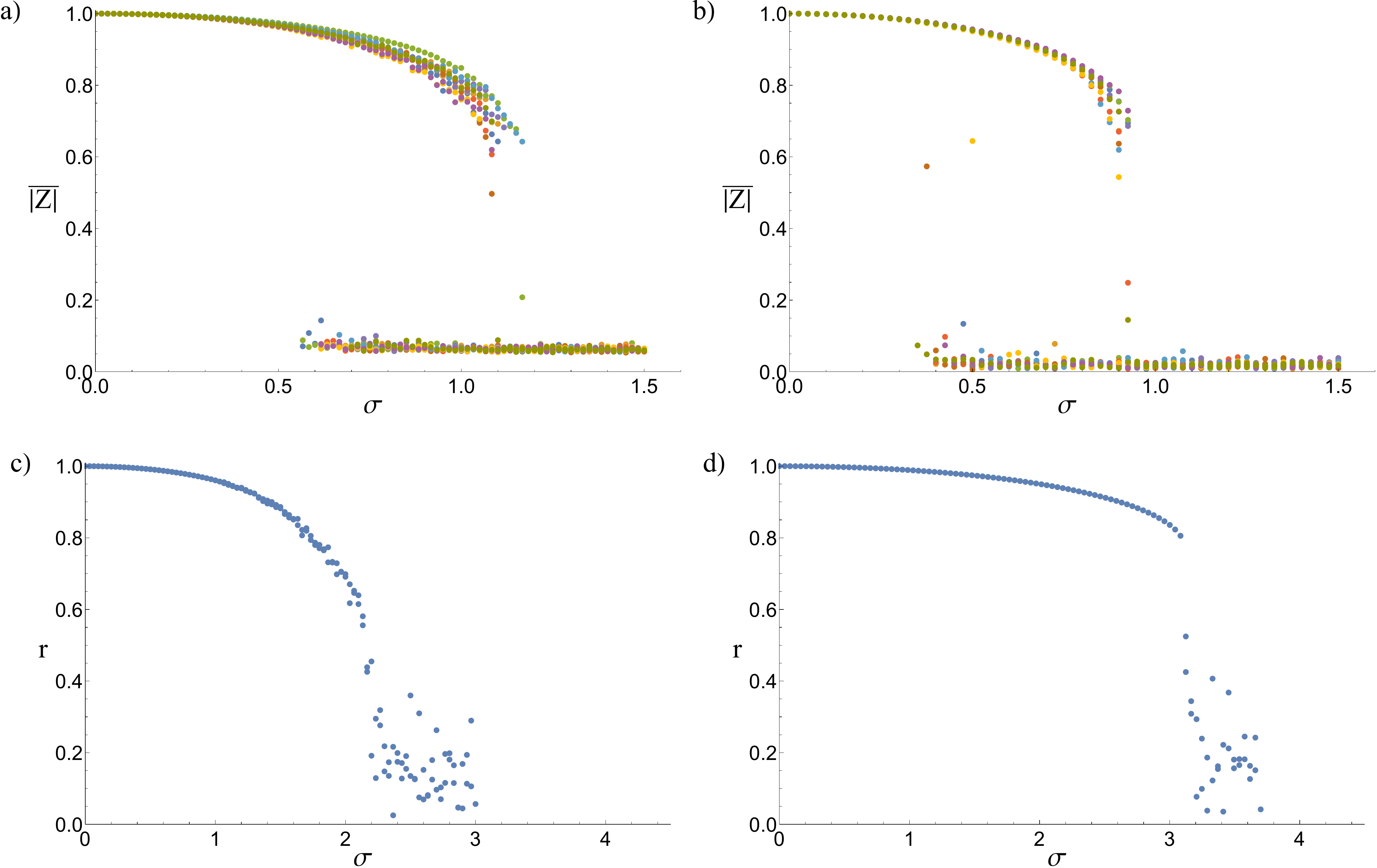}
\caption{Order parameter versus standard deviation $\sigma$ of disorder distribution $g(\omega)$. a) Ten realizations of Gaussian disorder in the ternary phase model for $N=300$, $K=4$. b) Ten realizations of uniform disorder in the ternary phase model for $N=300$, $K=2$. c) Gaussian disorder in the Kuramoto model for $N=400$, $K=4$. d) Uniform disorder in the Kuramoto model for $N=300$, $K=4$.}
\label{hysteresis}
\end{figure*}

Figures \ref{hysteresis}a) and \ref{hysteresis}b)
show typical results for random Gaussian and uniform disorder, respectively. The results are qualitatively similar. For small, increasing $\sigma$, the order parameter gradually decreases. At the critical value $\sigma_1$, there is a first-order transition beyond which the order parameter becomes negligible. Then, decreasing $\sigma$ induces another first-order transition at $\sigma_2$. The two transitions occur at different values of $\sigma$, resulting in a hysteresis loop. As expected, the fluctuations grow near the transition points. 

This behavior is different than what one sees in the Kuramoto model. For comparison, results for the latter are shown in Fig. \ref{hysteresis}c) (Gaussian disorder) and Fig. \ref{hysteresis}d) (uniform disorder).  For uniformly distributed disorder, the Kuramoto model shows a first-order transition with no hysteresis \cite{pazo}. In the case of Gaussian distributed disorder, it shows a second-order transition. 

\begin{figure*}[h!]
\centering
\includegraphics[scale=.4]{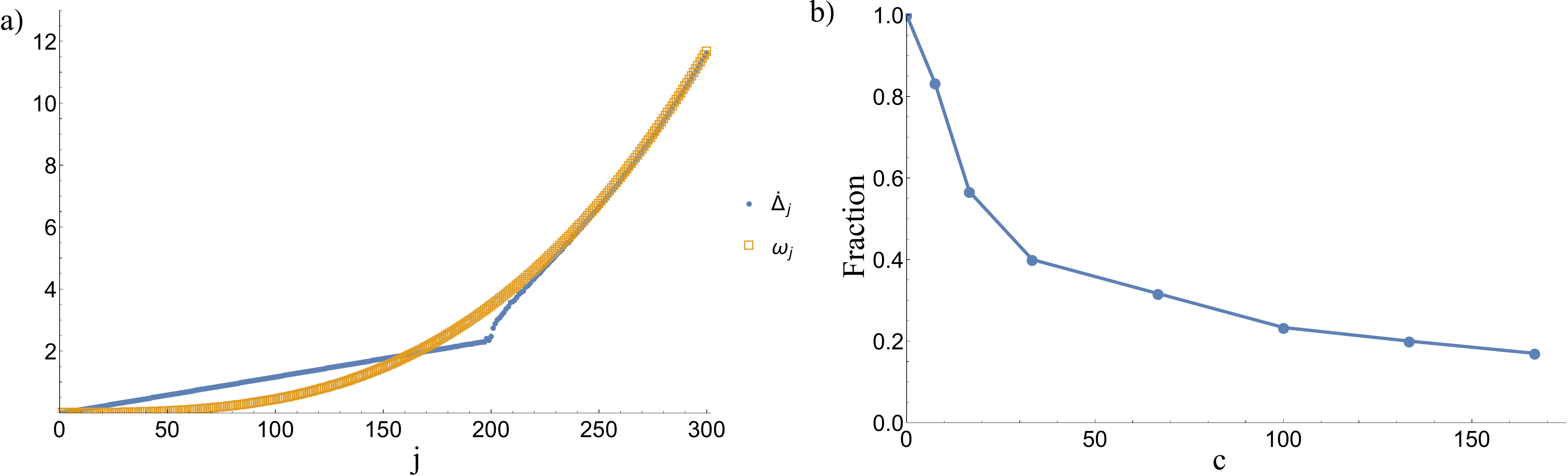}
\caption{Cubic disorder in the ternary phase model. a) $\dot{\Delta}_j$ and $\omega_j$ versus $j$ for the ternary phase model with cubic disorder at $c=12$. b) Fraction of locked oscillators versus width $c$ in the ternary phase model with cubic disorder. }
\label{cubic}
\end{figure*}

In addition to random disorder, we also considered disorder of the form
\begin{equation}
    \omega_j = \frac{c}{N^3} \cdot j^3
    \label{cubic_diffusion}
\end{equation}
with $c$ the effective ``width'' of the distribution. This choice is motivated by higher-order dispersion effects that arise in the micro-resonator problem \cite{taheri, wen}. The behavior in this case can be understood heuristically as follows. As $c$ is increased, the oscillators with large $j$,  i.e. those far from the pump mode, acquire an intrinsic frequency much larger than the coupling constant and thus become completely unlocked from the rest of the population. The lower $j$ oscillators still have relatively low disorder, and can synchronize. The result is that the population is synchronized up to a critical $j$ above which the oscillators are effectively free-running. This is illustrated in Figure \ref{cubic}a)
by plotting $\dot{\Delta}_j$ and $\omega_j$ versus $j$ in the non-co-rotating frame. The locked oscillators fall on a straight line reflecting the synchronization condition (\ref{tsync1}-\ref{tsync2}) while the unlocked oscillators, beginning at $j \approx 200$, rotate at their intrinsic frequencies. While the locked oscillators can be thought of as being frequency locked, this is really a result of their phase locking;  for the ternary model the two are necessarily connected through the synchronization condition.

Carrying out simulations using the same protocol as before, we find that the fraction of locked oscillators (those with $\dot{\Delta}_j$ falling on the straight line) decreases continuously to zero with no observed transition or hysteresis (Figure \ref{cubic}b). This suggests that cubic dispersion merely dilutes the total synchronization; for random noise, the first-order transitions enforce a critical threshold above which no synchronization is possible.

\section{Thermodynamic Limit}
In order to move beyond simulations, we formulated a thermodynamic limit corresponding to an infinite number of oscillators on a lattice with uniform spatial density. We show that, in this limit, there are no fluctuations and the collective state can be uniquely identified.

Unlike the all-to-all coupling of oscillators in the Kuramoto model, the ternary coupling in (\ref{ternary_with_disorder}) is index-dependent. We therefore consider the oscillators to lie on a discrete lattice corresponding to $j=1, 2, \dots ,N$. Inspired by Ref. (\onlinecite{gupta}), we consider a spatial continuum of oscillators which can be described by an explicitly position-dependent probability density.

First, we note that the equation of motion (\ref{ternary_with_disorder}) can be written in terms of the order parameters (\ref{orderp}) as
\begin{equation}
    \begin{split}
         \dot{\Delta}_j &= \omega_j + \frac{2N+1-j}{2N}KR_j\sin(\psi_j-\Delta_j), \\
         Z_j &= R_j \textrm{e}^{\textrm{i}\psi_j}.
         \label{meanfield}
    \end{split}
\end{equation}
In this form, the individual oscillators appear decoupled which allows for a self-consistent calculation of the order parameters. The order parameters $Z_j$ are written explicitly as 
\begin{equation}
    Z_j=\frac{1}{2N+1-j}\sum_{l=j-N}^{N}\textrm{e}^{\textrm{sgn}(l)\textrm{i}\Delta_{|l|}}\textrm{e}^{\textrm{sgn}(j-l)\textrm{i}\Delta_{|j-l|}}
    \label{orderpexp}
\end{equation}
reflecting the antisymmetry condition $\Delta_{-j} = -\Delta_j$. The $Z_j$ can be expressed in terms of a probability distribution rather than an explicit sum as
\begin{equation}
    \begin{split}
        Z_j=\frac{1}{2N+1-j}\sum_{l=j-N}^{N}\int\limits_0^{2\pi}\textrm{d}\Delta\int\limits_0^{2\pi}\textrm{d}\Delta'\int\limits_{-\infty}^\infty\textrm{d}\omega\int\limits_{-\infty}^\infty\textrm{d}\omega' \\
        \times P(\Delta, \Delta', \omega, \omega', |l|, |j-l|)\textrm{e}^{\textrm{sgn}(l)\textrm{i}\Delta}\textrm{e}^{\textrm{sgn}(j-l)\textrm{i}\Delta'}
    \end{split}
\end{equation}
where $P(\Delta, \Delta', \omega, \omega', x, x')$ is the joint probability density of the oscillator at $x$ having phase $\Delta$ and disorder $\omega$ while the oscillator at $x'$ has phase $\Delta'$ and disorder $\omega'$. In the spirit of Eq.(\ref{meanfield}), we consider the oscillators to be independent so that the probability density $P$ can be approximated as
\begin{equation}
    P(\Delta, \Delta', \omega, \omega', |l|, |j-l|)=\rho(\Delta, \omega, |l|)\rho(\Delta', \omega', |j-l|)
\end{equation}
with $\rho$ the probability density for a single oscillator parameterized by position.

Because the coupling depends on oscillators at two different positions, the density $\rho$ appears twice. This is one of the major structural novelties of the model and significantly complicates stability analysis of the unsynchronized state. 

In order to treat the infinite oscillator limit, in which case the probability densities vary continuously in space, we introduce a spatial coordinate $x$ such that
\begin{equation}
    \begin{split}
        x = \frac{j-1}{N-1}, \\
        0\leq x \leq 1.
    \end{split}
    \label{coords}
\end{equation}

From (\ref{coords}), $\textrm{d}j = (N-1)\textrm{d}x$, and the summation over $l$ transforms according to
\begin{equation}
    \frac{1}{2N+1-j}\sum_{l=j-N}^N \textrm{d}j= \frac{N-1}{2N-(N-1)x}\sum_{y=\frac{(N-1)x-N}{N-1}}^1 \textrm{d}y.
\end{equation}
Taking $N \rightarrow \infty$, the sum becomes an integral and position $x$ becomes continuous, leading to
\begin{equation}
    \begin{split}
        Z(x)=\frac{1}{2-x}\int\limits_{x-1}^1\textrm{d}y\int\limits_0^{2\pi}\textrm{d}\Delta\int\limits_0^{2\pi}\textrm{d}\Delta'\int\limits_{-\infty}^\infty\textrm{d}\omega\int\limits_{-\infty}^\infty\textrm{d}\omega' \\
        \times \rho(\Delta, \omega, |y|)\rho(\Delta', \omega', |x-y|)\textrm{e}^{\textrm{sgn}(y)\textrm{i}\Delta}\textrm{e}^{\textrm{sgn}(x-y)\textrm{i}\Delta'}.
    \end{split}
    \label{orderx}
\end{equation}
The full thermodynamic system, completely described by $\rho$, is governed by the continuity equation
\begin{equation}
    \begin{split}
        &\frac{\partial \rho}{\partial t} +     \frac{\partial}{\partial     \Delta}\big(\rho \dot{\Delta} \big)     = 0, \\
        \dot{\Delta} = \omega +&(1-x/2)KR(x) \sin(\psi(x)-\Delta), \\
        &Z(x) = R(x)\textrm{e}^{\textrm{i}\psi(x)}
    \end{split}
    \label{continuity}
\end{equation}
and the constraints
\begin{equation}
    \begin{split}
        &\int\limits_0^{2\pi}\textrm{d}\Delta\int\limits_{-\infty}^\infty \textrm{d}\omega \rho(\Delta, \omega, x) = 1, \\
        &\int\limits_0^{2\pi}\textrm{d}\Delta \rho(\Delta, \omega, x) = g(\omega)
    \end{split}
\end{equation}
where $g(\omega)$ is the probability density for each oscillator's disorder.

Having established the evolution equations for the thermodynamic limit, we may calculate the order parameter $R=\overline{|Z|}$ as a functional of the disorder distribution $g(\omega)$. The details of the calculation are relegated to the appendix, but the final result is 
\begin{equation}
    R = \int\limits_{0}^1\textrm{d}y\Bigg[\frac{(2-y)KR}{2}\int\limits_{-\frac{\pi}{2}}^{\frac{\pi}{2}}\textrm{d}\theta \; g\bigg[\frac{(2-y)KR}{2}\sin\theta\bigg] \cos^2\theta\Bigg]^2,
    \label{rcalc}
\end{equation}
valid for even, unimodal $g(\omega)$.  

\begin{figure*}[h!]
\centering
\includegraphics[scale=.4]{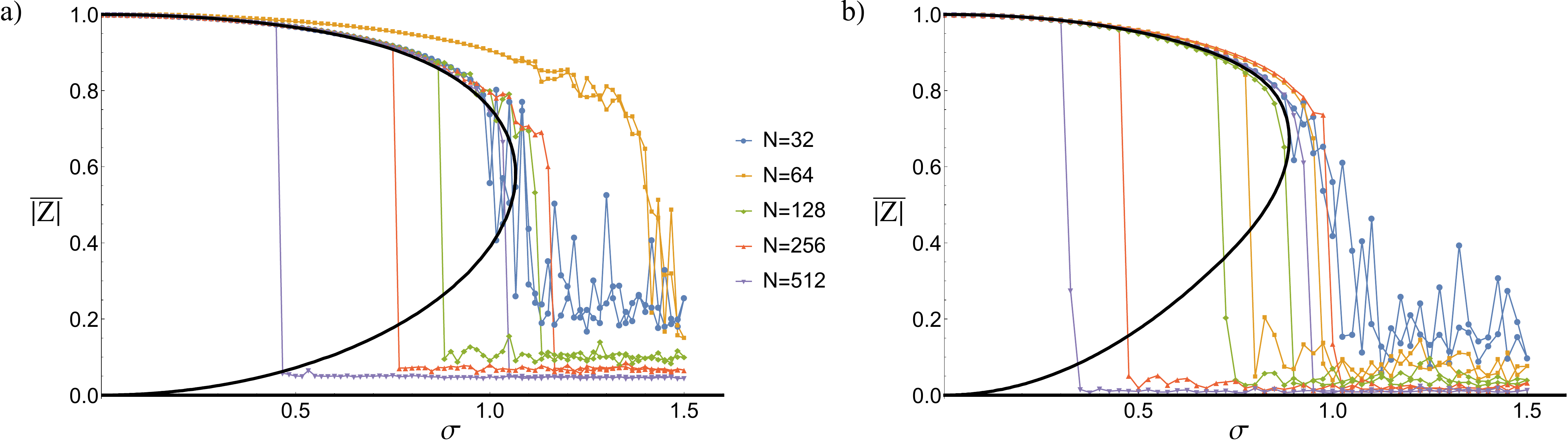}
\caption{Predicted order parameter in the infinite oscillator limit (thick line) for a) Gaussian disorder, b) uniform disorder. Agreement improves with a larger number of oscillators $N$.}
\label{prediction}
\end{figure*}

One sees immediately that $R=0$ is always a solution.  To find non-zero solutions, Eq.(\ref{rcalc}) can be numerically solved for any given $g(\omega)$. The resulting predictions for Gaussian and uniform disorder are shown in Figure \ref{prediction}
alongside values from direct simulation for different $N$. As $N$ is increased, agreement along the upper branch improves considerably. Since $|Z|$ is non-negative, convergence along the lower incoherent branch is much slower. The middle solution branch is presumably unstable, as is the case for a saturating subcritical pitchfork bifurcation \cite{strogatzbook}.

Curiously, the unsynchronized state is stable for a wider range of $\sigma$ as $N$ is increased, and may even be stable for all values as $N \rightarrow \infty$. This would be consistent with the theoretically predicted curve in which the middle branch and $R=0$ branch meet exactly at the origin. Bistability of the synchronized and unsynchronized states would have important implications for the micro-resonator problem since even a perfect device could fail to achieve mode locking.

 While our previous analysis hasn't addressed stability, we note that linear stability of even the trivial $R=0$ state is complicated by the fact that the order parameter is quadratic in $\rho$. This can be understood from the following argument without a complete stability calculation. The incoherent $R=0$ state is represented by the uniform density 
 \begin{equation}
     \rho_0= \frac{g(\omega)}{2\pi}.
 \end{equation}
 Evaluation of the order parameter (\ref{orderx}) using a perturbed solution of the form $\rho_0+\epsilon \delta\rho$ yields zero to linear order in $\epsilon$; one of the two factors of $\rho$ must be $\rho_0$, causing its associated exponential in $\Delta$ to vanish upon integration. However, the order parameter is responsible for coupling the oscillators as shown in (\ref{continuity}). When it vanishes, $K$ disappears from the equations and so cannot be responsible for a bifurcation. As a result, a nonlinear stability analysis is required to describe any synchronizing transition from the incoherent state.

\section{Conclusion and Open Questions}
In this paper, we investigated the synchronization behavior of a new phase model. We found that its transitions are characteristically different from those found in the classic Kuramoto model. We developed a thermodynamic, continuum limit which we used to calculate the order parameter for different disorder distributions. Although the predicted values for the order parameter agree well with simulations, the onset of instability for the unsynchronized solution is currently unexplained. With the enormous utility of the Ott-Antonsen ansatz\cite{ott} for the continuum limit of the Kuramoto model, it is natural to wonder if a similar dimensional reduction can be performed for the ternary model. 

In addition to its novel dynamical properties,  investigating the ternary phase model provides insight into the robustness of the synchronization process in optical microresonator cavities. Additional effects could be explored such as finite-size fluctuations and the inclusion of non-identical coupling constants $K_j$. The latter has been investigated in the absence of disorder \cite{taheri}. 

Perhaps the greatest outstanding challenge in studying the ternary phase model is the issue of stability. The difficulty of even linear stability analysis for the Kuramoto model is well known\cite{crawford}. For the ternary model, linear analysis fails and a strictly nonlinear approach is apparently needed. 

\appendix
\section{Calculation of the Order Parameter}
Two key observations from the discrete formulation allow us to proceed in calculating the thermodynamic limit of the order parameter for a given disorder distribution. First, the steady-state phase of the order parameter is linear in space, i.e. $\arg(Z_j) = \alpha j$. Since $\alpha$ is an arbitrary constant, it can be set to zero. Second, $|Z_j|$ is observed to be independent of $j$ in steady state. We can therefore choose to evaluate $Z(x)$ arbitrarily at $x=0$ and set its phase to zero. Equation (\ref{meanfield}) then simplifies to
\begin{equation}
    \begin{split}
        \dot{\Delta} &= \omega - M(x)\sin\Delta, \\
        M(x) &= (1-x/2)KR
    \end{split}
    \label{ddotx}
\end{equation}
where $R=|Z(0)|$. 

In the spirit of Kuramoto's solution to the Kuramoto model\cite{wiener, kuramotobook}, we seek stationary solutions of (\ref{continuity}) assuming a constant value of $R$. Oscillators with small disorder will have their phase entrained to the equilibria of (\ref{ddotx}) such that
\begin{equation}
    \begin{split}
        & \rho(\Delta, \omega, x) = \delta\bigg(\Delta-\arcsin\frac{\omega}{M(x)}\bigg)g(\omega), \\
        & \quad |\omega | \leq M(x).
    \end{split}
    \label{syncdens}
\end{equation}
The remaining oscillators are unsynchronized, and stationarity requires 
\begin{equation}
    \begin{split}
        & \rho(\Delta, \omega, x) = \frac{a}{|\dot{\Delta}|}g(\omega)=\frac{a}{\big|\omega - M(x)\sin\Delta\big|}g(\omega),\\
        & \quad |\omega | > M(x).
    \end{split}
    \label{unsyncdens}
\end{equation}
with $a$ determined by normalization.

Combining Eqs. (\ref{orderx}), (\ref{syncdens}), and (\ref{unsyncdens}) yields a self-consistency equation for the order parameter:
\begin{equation}
    \begin{split}
        R = &I_1+I_2, \\
        I_1 = & \frac{1}{2}\int\limits_{-1}^1\textrm{d}y\int\limits_0^{2\pi}  \textrm{d}\Delta\int\limits_0^{2\pi}\textrm{d}\Delta'\int\limits_{-M(|y|)}^{M(|y|)}\textrm{d}\omega\int\limits_{-M(|y|)}^{M(|y|)}\textrm{d}\omega' g(\omega)g(\omega')  \\
        & \times \textrm{e}^{\textrm{sgn}(y)\textrm{i} \Delta}\textrm{e}^{-\textrm{sgn}(y)\textrm{i} \Delta'} \\ & \times \delta\bigg(\Delta-\arcsin\frac{\omega}{M(|y|)}\bigg)\delta\bigg(\Delta'-\arcsin\frac{\omega'}{M(|y|)}\bigg), \\
        I_2 = & \frac{1}{2}\int\limits_{-1}^1\textrm{d}y\int\limits_0^{2\pi}  \textrm{d}\Delta\int\limits_0^{2\pi}\textrm{d}\Delta'\int\limits_{|\omega|>M(|y|)}\textrm{d}\omega\int\limits_{|\omega'|>M(|y|)}\textrm{d}\omega' g(\omega)g(\omega')\\
        & \times \textrm{e}^{\textrm{sgn}(y)\textrm{i} \Delta}\textrm{e}^{-\textrm{sgn}(y)\textrm{i} \Delta'} \\
        & \times\frac{a}{\big|\omega - M(|y|) \sin\Delta\big|}\cdot\frac{a}{\big|\omega' - M(|y|)\sin\Delta'\big|}.
    \end{split}
    \label{i1i2}
\end{equation}
For even, unimodal $g(\omega)$, (\ref{unsyncdens}) admits a symmetry in $\Delta$ and $\omega$ that causes the integrals over $\omega$ to vanish and so the unlocked oscillators do not contribute to the order parameter (see for example Ref.\onlinecite{wiener}). Evaluating the $\Delta$ integrals, (\ref{i1i2}) reduces to
\begin{equation}
    \begin{split}
    R = &\frac{1}{2}\int\limits_{-1}^1\textrm{d}y\int\limits_{-M(|y|)}^{M(|y|)}\textrm{d}\omega\int\limits_{-M(|y|)}^{M(|y|)}\textrm{d}\omega' g(\omega)g(\omega')\\
        & \times \exp\bigg[\textrm{i}\; \textrm{sgn}(y) \arcsin\frac{\omega}{M(|y|)}\bigg] \\ 
        & \times \exp\bigg[-\textrm{i}\; \textrm{sgn}(y) \arcsin\frac{\omega'}{M(|y|)}\bigg].
    \end{split}
    \label{nounsync}
\end{equation}
Expanding the complex exponentials, the imaginary terms are linear in $\omega$ and $\omega'$ respectively and thus vanish upon integration with even $g$. As a result, both integrands are even in $y$ and (\ref{nounsync}) reduces to
\begin{equation}
    \begin{split}
    R = \int\limits_{0}^1\textrm{d}y\Bigg[\int\limits_{-M(y)}^{M(y)}\textrm{d}\omega \; g(\omega) \sqrt{1-\bigg(\frac{\omega}{M(y)}\bigg)^2}\Bigg]^2.
    \end{split}
\end{equation}
Taking a change of variables
\begin{equation}
        \omega = M(y)\sin\theta,
\end{equation}
we end up with the indicated result
\begin{equation}
    \begin{split}
        R &= \int\limits_{0}^1\textrm{d}y\Bigg[M(y)\int\limits_{-\frac{\pi}{2}}^{\frac{\pi}{2}}\textrm{d}\theta \; g\big(M(y)\sin\theta\big) \cos^2\theta\Bigg]^2. \\
        &= \int\limits_{0}^1\textrm{d}y\Bigg[\frac{(2-y)KR}{2}\int\limits_{-\frac{\pi}{2}}^{\frac{\pi}{2}}\textrm{d}\theta \; g\bigg[\frac{(2-y)KR}{2}\sin\theta\bigg] \cos^2\theta\Bigg]^2
    \end{split}
\end{equation}

\bibliography{aipsamp}

\end{document}